\documentclass{article}
\usepackage[utf8]{inputenc}

\usepackage[english]{babel}
\usepackage[T1]{fontenc}
\usepackage{textcomp}
\usepackage[utf8]{inputenc}    

\usepackage[a4paper,top=3cm,bottom=2cm,left=3cm,right=3cm,marginparwidth=1.75cm]{geometry}

\usepackage{amsmath}
\usepackage{graphicx}
\usepackage[colorinlistoftodos]{todonotes}
\usepackage[colorlinks=true, allcolors=blue]{hyperref}
\usepackage{svg}
\usepackage{subfig}
\usepackage[separate-uncertainty = true,multi-part-units = repeat]{siunitx}
\usepackage{graphicx,rotating,booktabs}
\usepackage{multirow}

\usepackage{authblk}

\makeatletter
\newcommand{\settitle}{\@maketitle}
\makeatother

\title{RHEOS.jl -- A Julia Package for Rheology Data Analysis}
\author[]{J L Kaplan \thanks{J L Kaplan, email: jlk49@cam.ac.uk}}
\author[]{A Bonfanti \thanks{A Bonfanti, email: ab2425@cam.ac.uk}}
\author[]{A Kabla \thanks{A Kabla, email: ajk61@cam.ac.uk}}
\affil[]{Engineering Department, Cambridge University, UK}
\date{}

\begin{document}
\maketitle

\section*{Summary}

Rheology is the science of deformation and flow, with a focus on materials that do not exhibit simple linear elastic or viscous Newtonian behaviours. Rheology plays an important role in the characterisation of soft viscoelastic materials commonly found in the food and cosmetics industries, as well as in biology and bioengineering. Empirical and theoretical approaches are commonly used to identify and quantify material behaviours based on experimental data.

RHEOS (RHEology, Open-Source) is a software package designed to make the analysis of rheological data simpler, faster, and more reproducible. RHEOS is currently limited to the broad family of linear viscoelastic models. A particular strength of the library is its ability to handle rheological models containing fractional derivatives, which have demonstrable utility for the modelling of biological materials \cite{aime2018,bouzix2018,bonfanti2019, kaplan2019}, but have hitherto remained in relative obscurity -- possibly due to their mathematical and computational complexity. RHEOS is written in Julia \cite{bezanson2017}, which provides excellent computational efficiency and approachable syntax. RHEOS is fully documented and has extensive testing coverage.

To our knowledge, there is to this date no other software package that offers RHEOS' broad selection of rheology analysis tools and extensive library of both traditional and fractional models. It has been used to process data and validate a model in \cite{bonfanti2019}, and is currently in use for several ongoing projects.

It should be noted that RHEOS is not an optimisation package. It builds on another optimisation package, NLopt \cite{johnsonNL}, by adding a large number of abstractions and functionality specific to the exploration of viscoelastic data.

\section*{Statement of Need}

\subsection*{Arbitrary stress-strain curves and broad relaxation spectra require advanced software}

Many scientists and engineers who undertake rheological experiments would fit their data with one or several viscoelastic models in order to classify materials, quantify their behaviour, and predict their response to external perturbations.

Standard linear viscoelastic models take the form of an ordinary differential equation between stress $\sigma$ and strain $\epsilon$. Under simple perturbations (step or ramp in stress or strain, or frequency sweep), it is relatively straight-forward to extract time-scales and identify asymptotic behaviours required to identify parameter values. However, data often involves complex stress and strain signals, and materials whose behaviour involves a broad distribution of time-scales, including power law behaviours. Fitting models and predicting their response in the time domain then requires computing viscoelastic hereditary integrals such as:

$$ \sigma(t) = \int_{0}^t G(t - \tau) \frac{d \epsilon(\tau)}{d \tau} d \tau $$

where $G$ is the relaxation response (stress response to a step in strain) of the material. $G$ is defined analytically for classical models, but may only be available numerically in some cases. A similar relation exists to calculate the strain from the stress history. Fitting and predicting behaviour then becomes non-trivial and standardised processing tools are needed.

\subsection*{Learning about rheology is facilitated by the ability to explore a large database of models}

Obtaining intuition for fractional viscoelastic theory can be difficult and learning material is sparse: of popular rheology textbooks published over several decades \cite{barnes1989,findley1989,brinson2008,lakes2009,ward2013}, fractional viscoelasticity is only mentioned briefly in one of them \cite{lakes2009}. Tools are needed to support researchers with their exploration of standard and advanced models, and how they behave in response to idealised loading conditions, in particular when analytical expressions are difficult to obtain.

\subsection*{Extracting parameters, selecting models, and comparing materials require standardised tools}

Because understanding of materials is often dependent on summarising their behaviour with a model, one must be able to test and compare a broad range of models to inform model selection and reliably identify material parameters. There are currently very limited options available in the public domain \cite{bobrheology,seifert2019}, and most research groups have to invest significant effort into developing custom software. An open-source standardised library of models and fitting algorithms would support the rheology research community and make analysis more systematic, transparent, and reproducible.

\section*{Implementation}

\subsection*{Features}

RHEOS addresses the issues outlined in the Statement of Need in several ways.

\begin{itemize}
    \item RHEOS includes an extensive library of both traditional and fractional viscoelastic models. Although this library will satisfy most users, it is also straightforward to add additional models to RHEOS should they need to.
    \item As well as being able to fit models to experimental data and predict the materials response to step loading of stress or strain, RHEOS can handle arbitrary loading for non-singular and singular models, and for constant or variable sample rates.
    \item For intuition-building and model exploration, RHEOS includes signal generation features so that common loading patterns (e.g. step, ramp, stairs) can be applied to unfamiliar models.
    \item As a convenience to the user, RHEOS also includes easy-to-use CSV importing and exporting functions, as well as a number of preprocessing functions for resampling and smoothing.
\end{itemize}

All of the above features are linked together in a seamless interface intended to be very approachable for less experienced programmers. The different paradigms of creep, relaxation, and oscillatory testing are all accounted for, and models fitted against one type of data can be used to predict against a different type of data. (For instance, fitting against relaxation data and predicting the frequency response spectrum.)

\subsection*{Workflow}

\begin{figure}[t]
  \centering
 \includegraphics[width=0.8\textwidth]{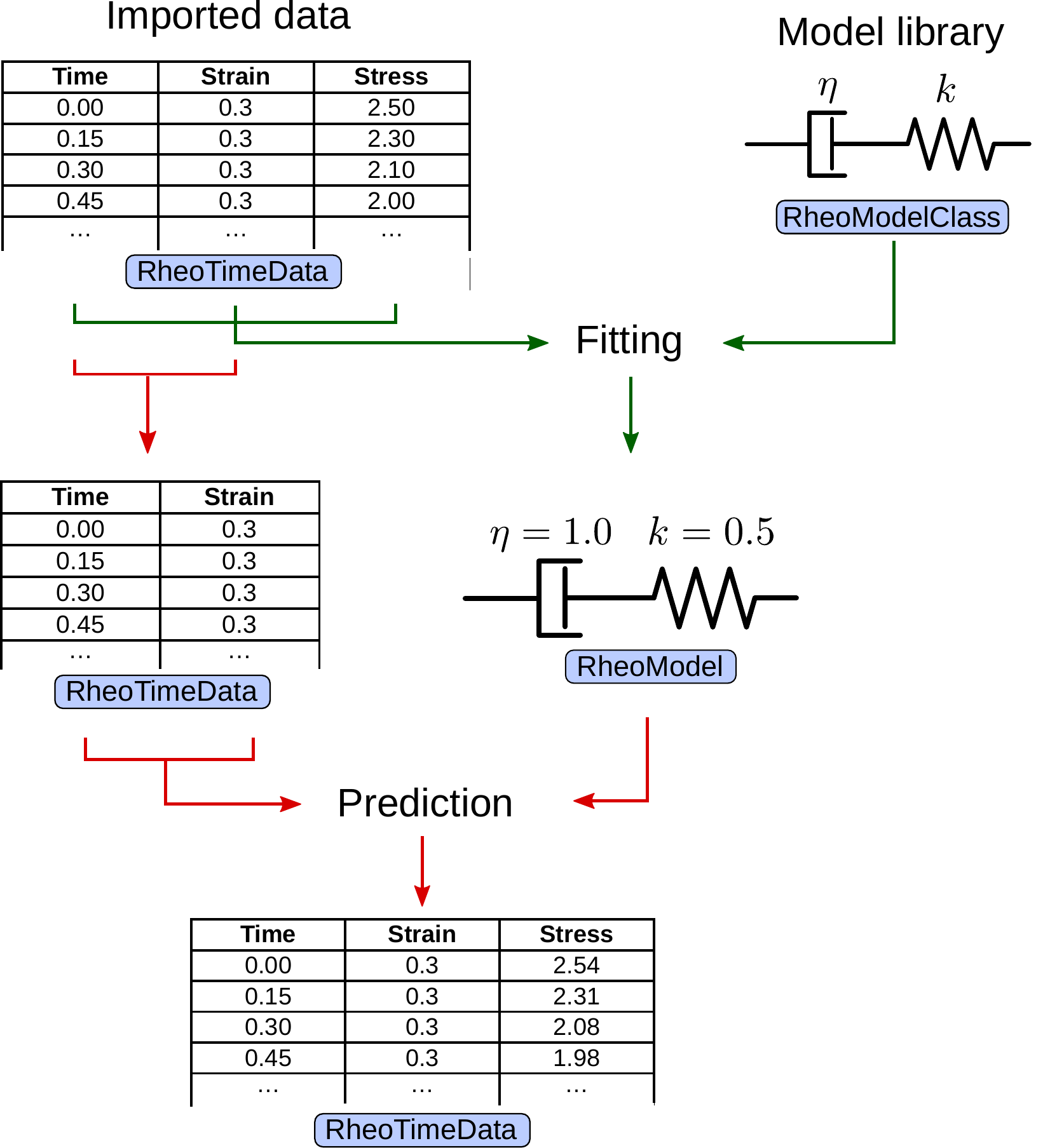}
  \caption{High level schematic of a fitting and prediction workflow from experimental data.}
  \label{fig:01}
\end{figure}

\begin{figure}[t]
  \centering
 \includegraphics[width=0.7\textwidth]{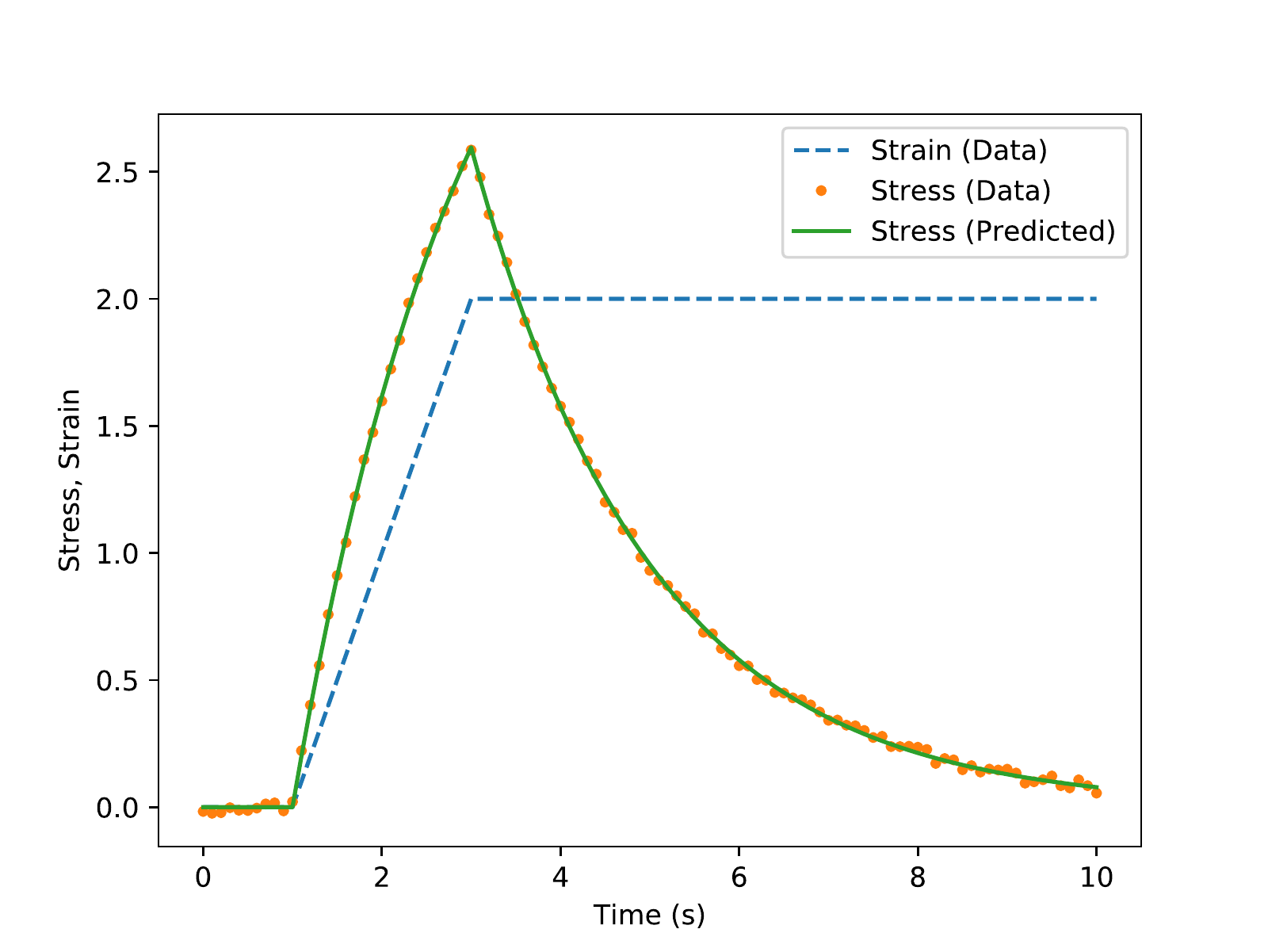}
  \caption{Qualitative assessment of the fitted model.}
  \label{fig:02}
\end{figure}

The following schematic illustrates one of the common RHEOS workflows in which experimental time-domain viscoelastic data is fitted to a model. This model is then used to make a prediction of the behaviour so that its accuracy can be qualitatively assessed. This workflow is shown schematically in Figure \ref{fig:01}, and the prediction of the fitted model is plotted against the original data in Figure \ref{fig:02}.

A brief description of this workflow is the following. A CSV is imported into a RHEOS `RheoTimeData' struct using a convenient loading function. This is then fitted to a `RheoModelClass', which embeds expressions for key characteristics of the model (relaxation function, creep response, complex modulus) involving symbolic parameters. This results in a fitted `RheoModel' where parameters are now substituted with fixed values derived from the fitting procedure. In the prediction step, the fitted `RheoModel' is combined with partial data (here only time and strain) to simulate the stress values expected from the model. The original data and model can then be compared graphically and numerically.

This example and others are available as Julia Jupyter notebooks, available in RHEOS GitHub repository and viewable on the RHEOS documentation website \cite{rheosgithub}.

\section*{Acknowledgements}

JLK would like thank the George and Lillian Schiff Foundation for the PhD funding which facilitated this project. AB, JLK, and AJK acknowledge the BBSRC grants BB/M002578/1, BB/K018175/1, and BB/P003184/1. All authors would like to thank Rohit Goswami, Grant Bruer, and Adam Beall for their valuable feedback during the review process.


\begin{thebibliography}{14}

\bibitem{aime2018}
Aime, S., Cipelletti, L., Ramos, L. (2018). Power law viscoelasticity of a fractal colloidal gel. Journal of Rheology, 62(6), 1429–1441. doi:10.1122/1.5025622


\bibitem{bouzix2018} 
Bouzid, M., Keshavarz, B., Geri, M., Divoux, T., Del Gado, E., McKinley, G. H. (2018). Computing the linear viscoelastic properties of soft gels using an optimally windowed chirp protocol. Journal of Rheology, 62(4), 1037–1050. doi:10.1122/1.5018715

\bibitem{bonfanti2019} 
Bonfanti, A., Fouchard, J., Khalilgharibi, N., Charras, G., Kabla, A. (2020). A unified rheological model for cells and cellularised materials. R. Soc. open sci.7190920, doi:10.1098/rsos.190920

\bibitem{kaplan2019} 
Kaplan, J. L., Torode, T. A., Daher, F. B., Braybrook, S. A. (2019). On pectin methylesterification: Implications for in vitro and in vivo viscoelasticity. bioRxiv. doi:10.1101/565614

\bibitem{bezanson2017}
Bezanson, J., Edelman, A., Karpinski, S., Shah, V. B. (2017). Julia: A fresh approach to numerical computing. SIAM Review, 59(1), 65–98. doi:10.1137/141000671

\bibitem{johnsonNL}
Johnson, S. G. The NLopt nonlinear-optimization package. Retrieved from https://github.com/stevengj/nlopt

\bibitem{barnes1989}
Barnes, H. A., Hutton, J. F., Walters, K. (1989). An introduction to rheology. Amsterdam, The Netherlands: Elsevier.

\bibitem{findley1989}
Findley, W. N., Lai, J. S., Onaran, K. (1989). Creep and relaxation of nonlinear viscoelastic materials with an introduction to linear viscoelasticity. Dover books on engineering. New York: Dover.

\bibitem{brinson2008}
Brinson, H. F., Brinson, L. C. (2008). Polymer engineering science and viscoelasticity: An introduction. New York: Springer.

\bibitem{lakes2009}
Lakes, R. (2009). Viscoelastic materials. Cambridge: Cambridge University Press. doi:10.1017/CBO9780511626722

\bibitem{ward2013}
Ward, I. M., Sweeney, J. (2013). Mechanical properties of solid polymers (Third edition.).Chichester, West Sussex, United Kingdom: Wiley.

\bibitem{bobrheology}
Das, C., Read, D. J., McLeish, T. C. (2012). Bob-rheology version 2.5. SourceForge. Retrieved from https://sourceforge.net/projects/bob-rheology/

\bibitem{seifert2019}
Seifert, J. (2019). Python tools for analysis of AFM data. Retrieved from https://github.com/jcbs/ForceMetric

\bibitem{rheosgithub} 
Kaplan, J. L., Bonfanti, A., Kabla, A. RHEOS.jl. GitHub repository https://github.com/JuliaRheology/RHEOS.jl(2020).


\end{thebibliography}
\end{document}